\documentclass[final,5p,times,twocolumn]{elsarticle}
\usepackage{lineno,hyperref}
\modulolinenumbers[5]

\usepackage{graphicx}
\usepackage{amssymb}
\usepackage{amsmath}
\usepackage{bm}
\usepackage{multirow}

\usepackage{amsmath}
\usepackage{subcaption}

\usepackage{xcolor}

\journal{}

\begin{document}

\begin{frontmatter}

\title{Pairwise interactions for Potential energy surfaces and Atomic forces with \\ Deep Neural network}


\author[1]{Van-Quyen Nguyen}
\author[2]{Viet-Cuong Nguyen}
\author[3]{Tien-Cuong Nguyen}
\author[5]{Nguyen-Xuan-Vu Nguyen}
\author[1,4]{Tien-Lam Pham\corref{mycorrespondingauthor}}

\address[1]{Phenikaa Institute for Advanced Study (PIAS), Phenikaa University, Ha Noi 12116, Viet Nam}
\address[2]{HPC SYSTEMS Inc.
Company Address: 3-9-15 Kaigan, Minato, Tokyo 108-0022, Japan}
\address[3]{VNU-University of Science, 334 Nguyen Trai, Thanh Xuan Dist., Hanoi, Vietnam}
\address[4]{Faculty of Computer Science, Phenikaa University, Yen Nghia, Ha Dong Dist., Hanoi 12116, Vietnam}
\address[5]{University of Science and Technology of Hanoi, Vietnam Academy of Science and Technology, 18 Hoang Quoc Viet, Hanoi, Vietnam.}

\cortext[mycorrespondingauthor]{Corresponding author}
\ead{lam.phamtien@phenikaa-uni.edu.vn}

\begin{abstract}
Molecular dynamics (MD) simulation, which is considered an important tool for studying physical and chemical processes at the atomic scale, requires accurate calculations of energies and forces. Although reliable energies and forces can be obtained by electronic structure calculations such as those based on density functional theory (DFT), this approach is computationally expensive. 
In this work, we propose a full-stack model using deep neural network (NN) to enhance the calculation of force and energy, in which the NN is designed to extract the embedding feature of pairwise interactions of an atom and its neighbors, which are aggregated to obtain its feature vector for predicting atomic force and potential energy.
By designing the features of the pairwise interactions, we can control the performance of models and take into account the many-body effects and other physics of the atomic interactions. 
Moreover, we demonstrated that using the Coulomb matrix of the local structures in complement to the pairwise information, we can improve the prediction of force and energy for silicon systems and the transferability of our models is confirmed to larger systems, with high accuracy. 

\end{abstract}


\end{frontmatter}

%\linenumbers

\section{Introduction}

	The computation of the energy, especially the forces of a chemical system, plays a central role in the computer simulation of matter and material design, which allows the sampling in the phase space by MD or Monte Carlo simulations. The energies and forces can be obtained by performing electronic structure calculations such as those based on density functional theory (DFT) \cite{HohenbergKohn1964,KohnSham1965}. 
	Although DFT has become a standard method in the computer simulations of matter \cite{CarParrinello1985}, these tools still have several limitations in versatility, veracity, and time-intensiveness.
	To overcome the issue of computational cost, the energies and forces of a large system can also be approximated by using an empirical model with a much lower computational cost, in which the potential energy surface (PES) is constructed as the accumulation of simple low-dimensional terms (structural elements) representing covalent bonds, angles, and dihedral angles \cite{Christopher}, and the force are calculated from the gradient of PES with respect to atomic coordinates. 
	These methods are efficient and are applied widely for simulating large biosystems, but they can hardly describe chemical reactions and other important atomic processes that involve the formation or dissociation of covalent bonds, and more accurate estimations are still highly necessary. 
	
	In recent years, alternative methods, which learn the PES from a set of materials structures and the corresponding DFT energies, have been developed \cite{BehlerPRL,Bartok_PRL,Bartok_tutorial,Bartok_SOAPs,SekoPRB,Rupps,JCP_LMM}. These methods employed machine learning algorithms to determine the functional relationship between the structure and energy of a material. Normally, to build the machine learning models the total energy of a material is decomposed into the sum of all energies of the constituent atoms determined by interactions among these atoms and with atoms in the surrounding chemical environment within a cutoff radius. Behler and colleagues \cite{BehlerPRL,BehlerJCP,Nongnuch_Artrith_nanoparticles,Eshet,Eshet1,Artrith,Artrith1} used atom-distribution-based symmetry functions to represent the local chemical environments of atoms and employed a multilayer NN to transfer the information of these chemical environments into the associated local (atomic) energies. This method is known as one of the most efficient methods for representing PES. Bart\'ok and coworkers \cite{Bartok_PRL,Bartok_tutorial,Bartok_SOAPs} employed the atomic density distribution to compare molecules and solids. Gaussian kernels were used to smoothly approximate the atomic density in a local structure. In addition, to develop machine learning models for the local energy, the similarity between two local structures was estimated by overlapping their atomic densities, which were expanded by spherical harmonic functions. These methods require the use of predefined basis functions to encode the primitive information of structural units such as two-body and three-body terms. 
	
	Although atomic forces, which is a vector quantity, can be calculated by obtaining the gradient of the total energy, this approach would hardly be successful when the learned functional relationship between the material structure and energy is complicated, such as deep-learning models with the sparsed representation of the structures. 
	Therefore, the machine learning models for representing the atomic forces in complement with energy can facilitate the MD calculation. 
	Further, in contrast to a large number of studies concerned with learning the atomic PES from data, a limited number of studies focusing on learning the atomic force have been reported thus far \cite{force_prediction_PRL,force_prediction_2}. Recently our studies \cite{JCP_LMM,doan,vnu_linear_regression}, as well as the work of Seko  \cite{SekoPRB} showed that the two-body term can be used to successfully represent the potential energy surfaces. 
	In this work, we present a model using deep NN by considering the two-body terms as the pairwise interactions in a certain chemical environment, in which the hidden features of the pairwise interaction are extracted from its features and its chemical environment, and the networks are shared by all pairs of atoms. 
	The embedding features of the pairwise interactions of an atom and its neighbors are then used to calculate  its feature vector for estimating its force and energy. For model predicting both force and energy of a system, we obtained root mean square error (RMSE) for the atomic force and energy of approximately $0.1048$ $eV/\AA$ and $4.531$ $meV/atom$, respectively. To improve the accuracy, we separated the model to two corresponding model that are responsible for each task. We demonstrated that the pairwise atomic forces derived by DFT can be well reproduced with a RMSE of approximately $0.0716$ $eV/\mbox{\AA}$ and $1.832$ $meV/atom$ for the silicon force and energy, respectively. 
	The physics of the interaction between atoms, i.e., the pairwise-force curves and basis functions for two-body terms can be deduced by analyzing the output of the hidden layers of the deep NN. 
	We revealed that our model has high transferability to apply to larger systems. Specifically, our models trained on 2x2x2 silicon systems can be utilized to 3x3x3 systems with RMSE for force and energy of $0.095$ $eV/\mbox{\AA}$ and $6.311$ $meV/atom$, respectively. Furthermore, we also examined our model for multi-component system. For Silicon Lithium system, we obtained the RMSE for force and energy of $0.118$ $eV/\mbox{\AA}$ and $4.117$ $meV/atom$, respectively
		
	\begin{figure}[!h]
		\centering
		\includegraphics[width=0.5\textwidth]{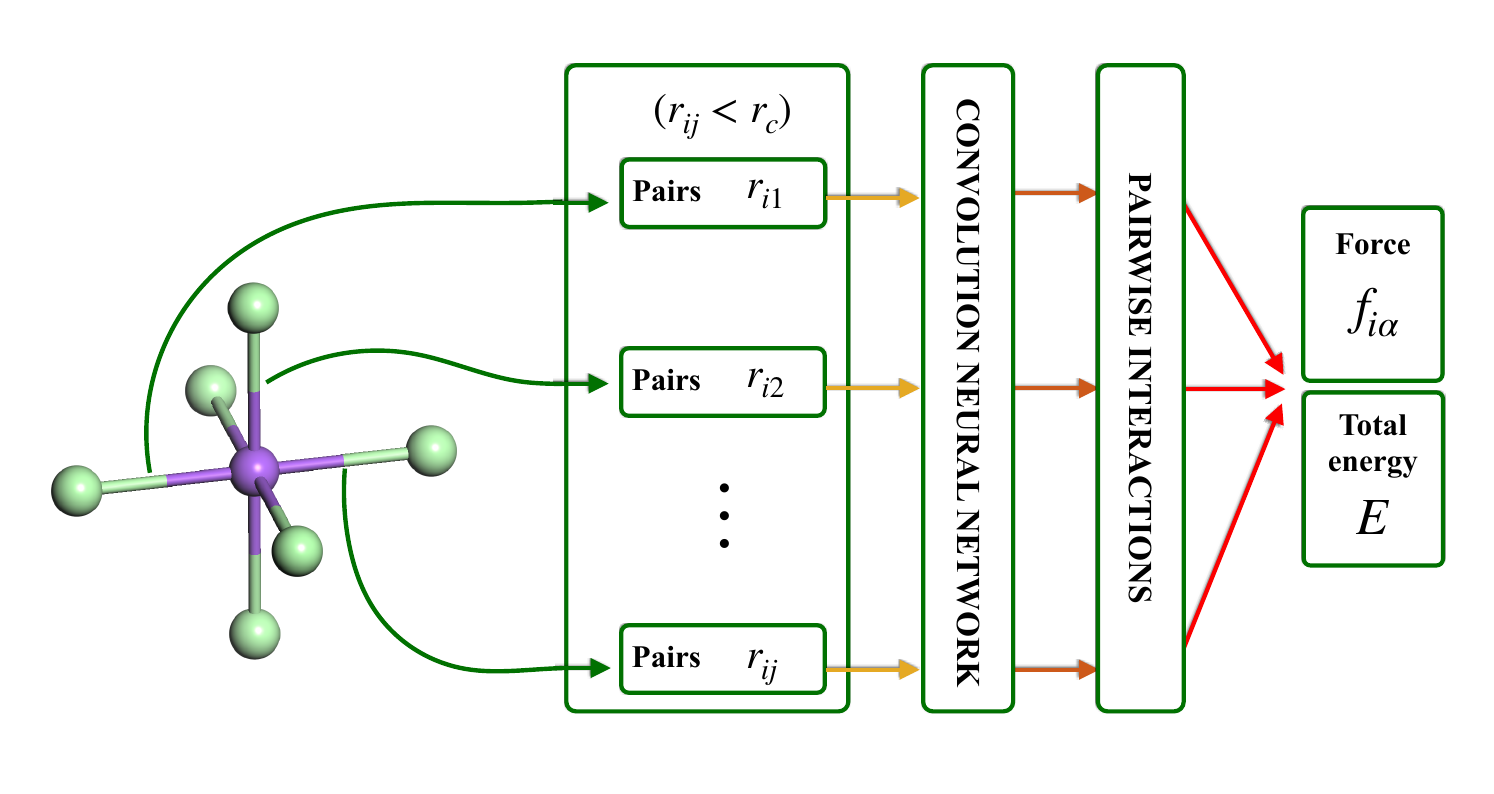}
		\caption{\label{fig:cnn_model}Model for prediction of energy and atomic forces: a local structure within a cutoff radius is decomposed into pairs of atoms (pairwise terms). The information of pairs terms is converted into embedding features and then transformed into pairwise force function or atomic energy for making the prediction of force and energy.}
	\end{figure}
	
	The remainder of this paper is organized as follows: The second section presents the formulation of models for force and energy; Section 3 illustrates experiment results and discussion; and the summary of the study is elucidated in the conclusion section.

\section{Methodology}

	One difficulty associated with utilizing machine learning in materials science is that the input to a predictive model for a property of a chemical system, can be of arbitrary size and shape depending on the size of studying systems. 
	To date, conventional machine learning algorithms have required input with a fixed-dimensional vector or matrix, and normally, the chemical systems have to convert into fixed-dimensional feature vectors, and during training, the feature vectors are also treated as fixed.  
	
	Moreover, it is well known that many properties of materials can be decomposed into the contributions of individual constituent structural elements such as chemical bonds, bond angles, and dihedral angles \cite{Christopher}. 
	For instance, the total energy of a chemical system can be expressed in terms of the bond potentials, bond-angle potentials, Lennard--Jones potential, etc., in the form of the classical potentials. 
	Another example used small structural fragments to represent the energy of a chemical system in the formulation of cluster expansion \cite{cluster_doi:10.1021/acs.jctc.8b00149}. 
	This consideration inspired us to develop a method to decompose a chemical structure into its structural elements, then design a machine learning model to extract hidden information of structural elements for estimating the material properties.
	In this study, we decompose a chemical structure into the collections of pairwise interactions in a certain chemical environment and adopt deep NN to extract embedding representation of the interaction for predicting atomic forces and total energy.  We design the network being able to apply for all the pairs of atoms, hence we can utilize it to systems of any size.

\subsection{Model for Energy by Pairwise interactions}
	The total energy of a solid system can be considered as the function of its atomic structure, i.e., the unit cell vectors and coordinates of atoms, and is invariant under the translations and rotations. To develop the models for representation of the total energy, ${E}$, we decompose it into the contributions constituent atoms, named atomic energy (partial energy) \cite{BehlerPRL,Zhang2018,JCP_LMM}: $E = \sum_i^{N}\bm{E}_i$, where $\bm{E}_i$ is the fictitious energy of $i^{th}$ atoms in the system and $N$ is the number of atoms in the system. The energy of an atom in the system is determined by its interaction with neighboring atoms (chemical environment) within a cutoff radius. By defining the center atom and its chemical environment as a local structure featured by the set of two-body terms, $\bm{x}_i$, the energy  can be expressed as follows: 
	\begin{equation}
		E = \sum_i^{N}\bm{E}_i = \sum_i^{N} F({\bm{x}_i}) 
	\end{equation}
	For instance, if we use one hidden layer the atomic energy can be calculated as follows: $\bm{E}_i = F({\bm{x}_i}) = {\bm{W}_2} \times g({\bm{W}_1} \times {\bm{x}_i})$, where ${\bm{W}_1}$ and ${\bm{W}_2}$ are the weights of the hidden layer and output layer, and $g$ is a nonlinear activation function. It notes that we use a single output layer with linear activation to obtain atomic energy, and the weights are shared by all local structures.  
	
	We derive feature vector, $\bm{x}_i$, of a local structure from the set pairwise terms represent the interaction of the center atom and its neighboring atoms, and we represent the pairwise term by feature vector: $\bm{b}_{ij}$ for the interaction of atom {$i$} and {$j$}. It notes that the number of neighboring atoms varies from local structure to local structure. We follow Behler \cite{BehlerJCP} method using symmetry basis functions, except that our basis functions are learned by neural network: $\bm{x}_{i} = \bm{x}_{i}(\{\bm{b}_{ij}\}) = \sum_j f(\bm{b}_{ij})$. It notes that $f(\bm{b}_{ij})$ is a vector function and maps the feature vector of the pairwise terms into embedding feature vectors with desired dimensions considered as basis functions, say $\bm{a}_{ij}$. This function is shared by all pairs of atoms, and we adopt a deep NN to represent it. In this work, we employ a neural network with three hidden layers, and the  embedding feature vector can be expressed as follows:  $\bm{a}_{ij} = {w}_3 \times g({w_2} \times g({w_1} \times \bm{b}_{ij}))$, where ${w_1}$, ${w_2}$, and ${w_3}$ are weights of hidden layers, and $g$ is a non-linear activation function.
	
	\begin{figure}[!h]
		\centering
		\includegraphics[width=0.45\textwidth]{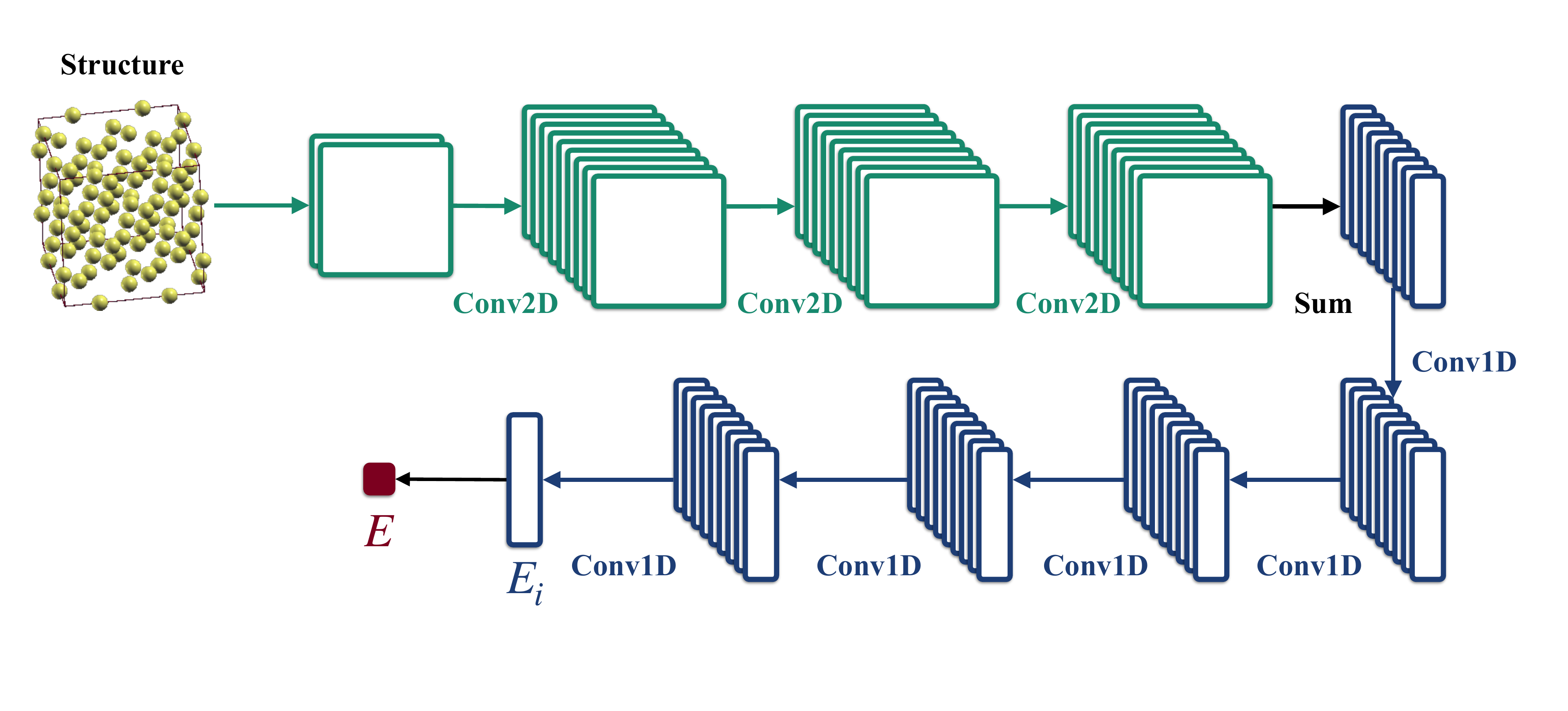}
		\caption{\label{fig:model_energy}The deep neural network model for representing the total energy: the structures are converted into 3D tensors of pairwise terms, then three Conv2D layers with a kernel size of $1 \times 1$ are applied the tensors to extract the embedding features, then the output of these layer are summed for representing pairwise interactions of an atom with its neighboring atoms. Finally, five Conv1D layers with the kernel size of $1$ are used for atomic energies.}
	\end{figure}

	To utilize the common deep learning framework, we define the representation vector of local structures as the descending-sorted by the distance to the center atoms of the set of their pairwise interaction terms of $\bm{b}_{ij}$: ${B} = (\bm{b}_{i1}, \bm{b}_{i2}, ..., \bm{b}_{in})$, where $n$ is the number of atoms in the chemical environment of atom $i$. Since the number of atoms in the chemical environments is not unique for the local structures, we employ the padding technique to ensure that all vectors have the same dimensions, i.e., append zero elements to the end of the shorter vectors. By this procedure, we can obtain the local structure by a matrix whose number of rows and columns being the maximum number of neighboring atoms and the dimensions of the pairwise terms. Finally, we stack the matrices of the local structures to make a 3D input tensor as the representation for a structure, in which the first dimension of the 3D input tensor is the number of atoms in the systems, the second dimension is the maximum number of neighboring atoms in chemical environments, and the third dimension is the number of dimensions of pairwise vectors. We employed the convolutional networks to implement our model, as seen in Fig. \ref{fig:model_energy} in which 2D Convolutional and 1D Convolutional layers were applied to extract hidden basis fuctions and atomic energy, respectively.
	
\subsection{Atomic force from Pairwise interactions}
	We consider the atomic forces that arise from their neighboring atoms within a cutoff radius. The force that acts on an atom is a superposition of all pairwise forces between the central atom and its neighboring atoms:
	\begin{equation}
		\label{eq:force}
		\bm{f}_i^\alpha = \sum_j \bm{f}_{ij}^\alpha 
		= \sum_j \bm{F}_{ij} \frac{\bm{r}_{ij}^\alpha}{|\bm{r}_{ij}|}
		= \sum_j \bm{F}_{ij}\frac{\bm{r}_{j}^\alpha - \bm{r}_{i}^\alpha}{|\bm{r}_{ij}|}f_c(\bm{r}_{ij})	
	\end{equation}
	where $\bm{f}_i^\alpha$ is the $\alpha$ component of force acting on atom $i$, $\bm{f}_{ij}^\alpha$ is the $\alpha$ component of the force exerted on atom $i$ by atom $j$, $\bm{F}_{ij}$ is a pairwise-force function, $\alpha \in \{x, y, z\}$, and $f_c(\bm{r}_{ij})$ is a smooth cutoff function.
	We use the cutoff smooth function proposed by Behler \cite{BehlerPRL,BehlerJCP}: $f_{c} = \frac{1}{2} \biggl{[} \biggl{(}cos \frac{\pi \bm{r}_{ij}}{r_{c}}\biggl{)}+1\biggl{]}$. The pairwise-force function, $\bm{F}_{ij}$, measures the strength of the force atom $j$ exerts on atom $i$, and depends on the pairwise terms and their chemical environment. 

	For each atom, say the $i^{th}$ atom, we acquire its neighboring atoms (indexing by $j$) within a cutoff radius, and the feature vectors representing the pairwise interactions, $\bm{b}_{ij}$. It notes that although we mention here the pairwise interactions, by suitable definition of these feature vectors one can add many-body effect by considering the pairwise terms in a chemical environment. We also adopt multi-layer perceptrons to represent the pairwise force function, $\bm{F}_{ij} = F(\bm{b}_{ij})$, and the networks are shared by all pairwise terms. For instance, if once employ the network with a hidden layer and an output neuron for $\bm{F}_{ij}$, then we can express the pairwise force as follows: $\bm{F}_{ij} = {w}_2 \times g({w}_1 \times \bm{b}_{ij})$.
	\begin{figure}[!h]
		\centering
		\includegraphics[width=0.45\textwidth]{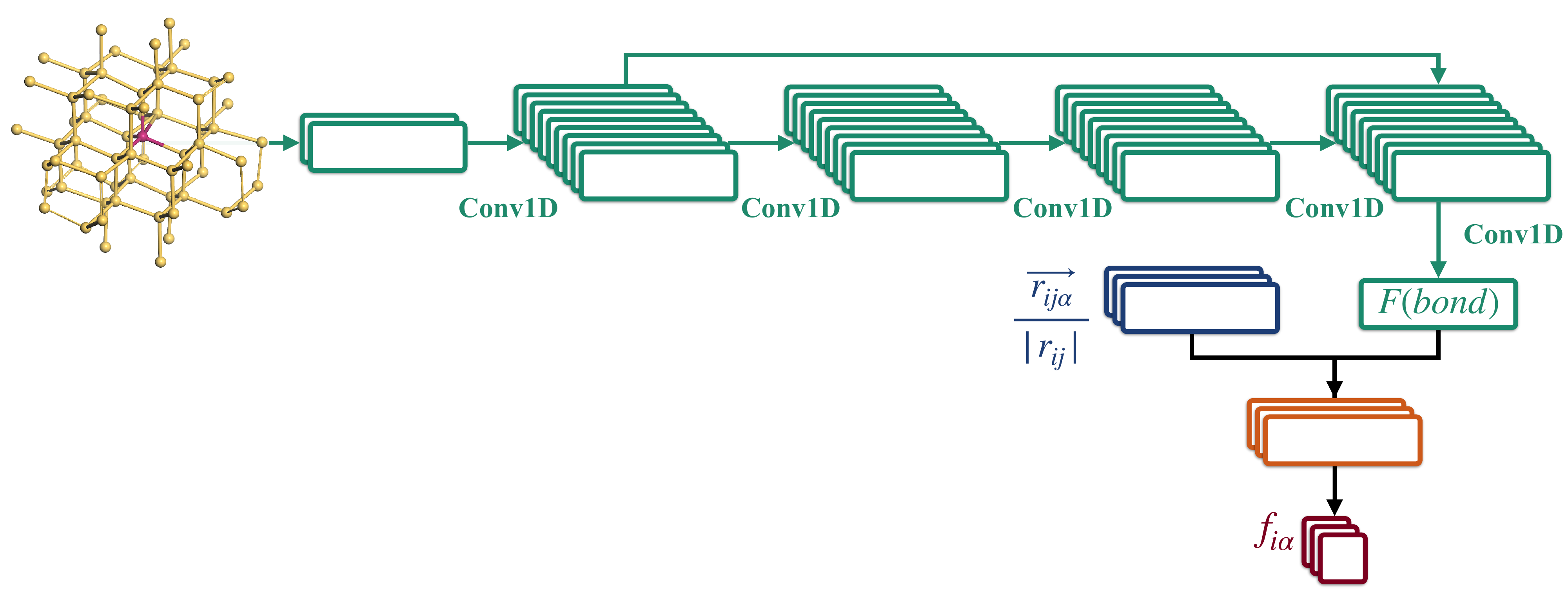}
		\caption{\label{fig:model_force}Convolutional Neural Network architecture for atomic forces: a local structure including the center atom and its neighboring is converted into a $n \times 2$ matrix, then five Conv1D layers with the kernel size of $1$ are applied to extract hidden pairwise force functions, the residual connection is added for the first layer and the last layer, the Conv1D with one neuron is used for the pairwise force function, and finally, the unit vector is multiplied to the pairwise force function to obtain atomic forces.}
	\end{figure}
	
	We also utilize the convolutional network technique to implement the force model. For each atom, we can derive the feature vectors for all pairwise interaction, $\bm{b}_{ij}$, within a cutoff radius and the unit vectors: $\frac{\bm{r}_{ij}}{|\bm{r}_{ij}|}$ for all pairs of atoms. We stack these vectors in the order of descending of the distance to the center atom and obtain the matrix of features, ${B}$, and matrix for unit vectors, ${N}$. We also utilize the padding technique to make the matrices representation of all atoms have the same dimensions. We now implement the Conv1D layers with the input of the matrix features, ${B}$, with the last layer we use only one neuron and the output of the Conv1D layers is a list of the pairwise forces. By multiplying these forces to corresponding unit vectors, we can obtain the forces of the neighboring atoms exerting on the central atom.   Finally, we use equation \ref{eq:force} to calculate the force on the atom of interest.
	
\subsection{Data}
	To examine our method, we employed our formulation to represent the energy and force of the silicon system. We trained our models using $2\times2\times2$ Si supercell that includes 64 Si atoms as both training and test data and a $3\times3\times3$ Si supercell with 216 atoms as the test data to estimate the transferability to the larger system. To make the reference data, we randomized the structure by adding random distortion to the cell vectors as well as atomic coordinates. To obtain the random distortion, three-dimensional deformation is applied with the magnitude of within 10 percent the lattice parameter in the $x$, $y$, and $z$ directions, and the strains are: $\Vec{\Gamma}' = (I + \varepsilon)(\Vec{\Gamma})$, where $\Vec{\Gamma'} = (a^{'}, b^{'}, c{'})$ is cell vectors of deformed structure, $\Vec{\Gamma} = (a, b, c)$ is cell vectors of the original (un-deformed) structure, $I$ is the identity matrix, $\varepsilon$ is random deformation parameters. We implemented the deformations and transformation to atomic positions and cell parameters to generate structure before calculated by DFT.

	We collected 5000 random structures for $2 \times 2 \times 2$ system and 500 structures for $3 \times 3 \times 3$ system.The energies and forces of the structures were calculated by using the PWSCF code \cite{pwscf_1,pwscf_2}. We employed the PBE exchange-correlation functional \cite{ggapbe1,ggapbe2} to process the exchange-correlation energy, plane-wave basis set with the cutoff energy of 40 Ry and ultrasoft-pseudopotential \cite{USPP} for approximating the interaction between the valence electrons and core electrons. A $3\times 3\times 3$ Monkhorst--Pack grid of k-points was employed for the Brillouin-zone integrations of $2\times 2\times 2$ Si supercell, while a $2\times 2\times 2$ grid was applied for $3\times 3\times 3$ supercell.

	We also examine the ability of the DNN model to multicomponent systems. We focus on the silicon lithium system (SiLi) which six Li atoms were randomly placed in a $2\times2\times2$ Si supercell that includes 64 Si atoms. We relaxed the system by employing an MD calculation and the Car-Parrinello method at 2,000 K in an NPT ensemble. The system was maintained at ambient atmospheric pressure (1.101325 bar) and then equilibrated at 1,000 K. We collected 5,000 structures in the trajectory as a multi-component system data set. The forces and energy are calculated by PWSCF code with the above setting.

\section{Result and Discussion}

\subsection{Potential Energy Model}

	To evaluate our deep model for total energy, we adopted the dataset of 5000 structures $2 \times 2 \times 2$  silicon systems with DFT-calculated energy. We divided the dataset into the training set of 3750 structures and the test set of 1250 structures, respectively. We employed the cutoff radius of 8.0 $\AA$ to determine the chemical environments for each atom. The partial energy associated with an atom was estimated by its interactions with neighboring atoms in the chemical environment. The interaction was considered as the pairwise interactions of the center atom and the neighboring atoms.  We designed the feature vectors for describing the pairwise interaction of atom $i$ and $j$ as follows: $\bm{b}_{ij} = (\bm{r}_{ij}  f_c(\bm{r}_{ij}), \frac{1}{\bm{r}_{ij}} f_c(\bm{r}_{ij}))$, where $r_{ij}$ is the distance of atom $i$ and  atom $j$. To obtain the embedding feature vector for the pairwise terms, we fed the feature vector to a three-layer perceptron network, and this network is shared by all pairwise terms. And the embedding feature vector for an atom is the sum of the output of the network of its pairwise interactions with neighboring atoms. We applied another three-layer perceptron network to convert the information in the embedding feature vectors of an atom into its partial energy (atomic energy), and this network is also shared by all atoms in the systems.

	To implement our idea, we designed the representation of an atom by stacking all pairwise vectors of its interactions with its neighboring atoms in the descending distance to the center atom. To ensure that all atoms were represented by the same dimensional matrices, we employed the padding technique.  In this particular silicon system, we used $150 \times 2$ matrices for all silicon atoms. We then stacked the representation matrix of all constituent atoms to make the descriptors for a structure as the 3D tensor, e.g., for this silicon system a structure is associated with $64 \times 150 \times 2$ tensors.  It is noted that this representation of the structure is similar to a color digital image, except that we only have two channels for the structure.
	
	\begin{figure}[!h]
		\centering
		\includegraphics[width=0.47\textwidth]{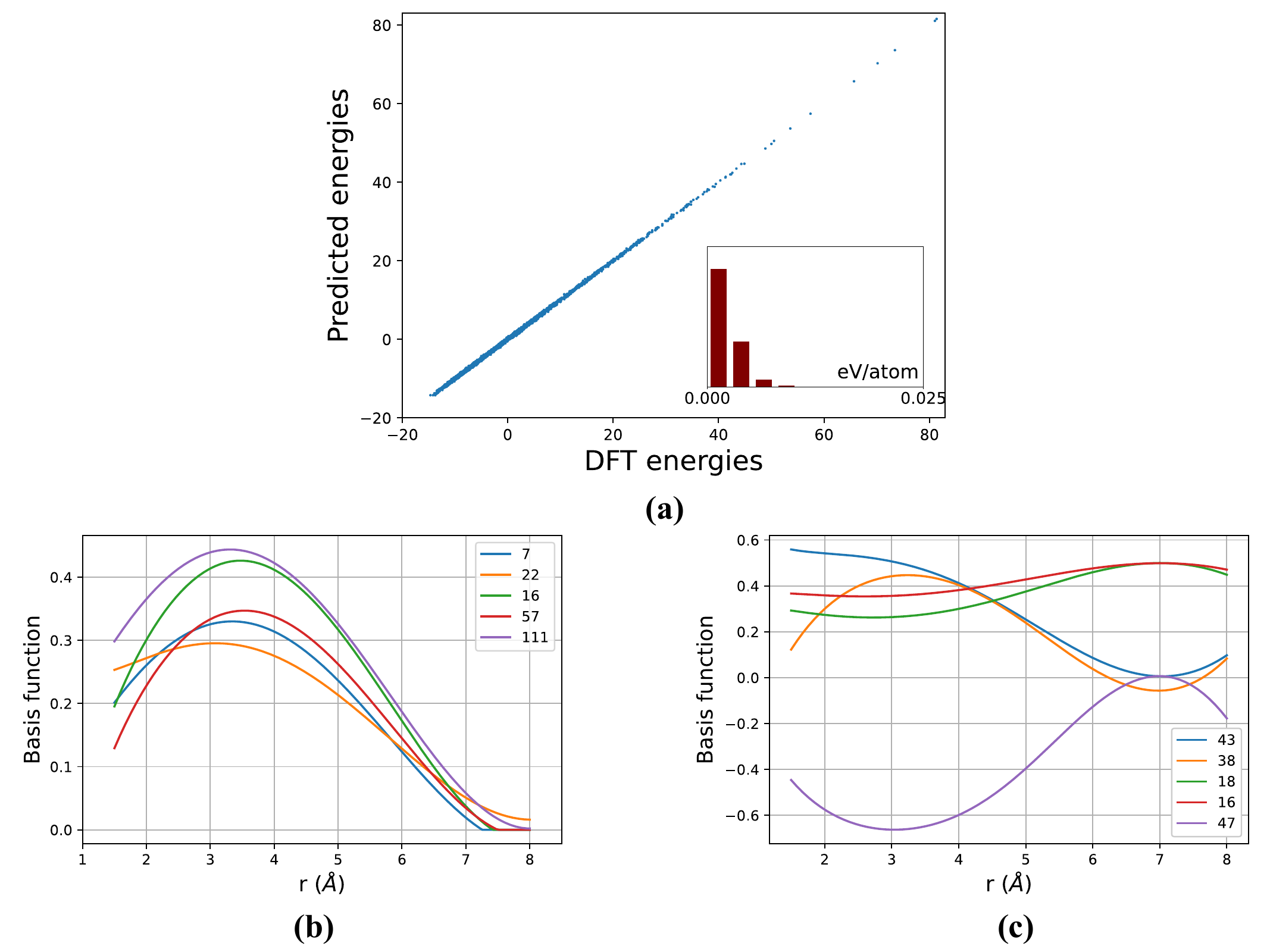}
		\caption{\label{fig:energy}Comparison of energy predicted by machine learning and DFT for $2 \times 2 \times 2$ system in the test set (a); basic functions of the pairwise terms derived by Relu activation (b) and mix activation (c): labeled by basis function index.}
	\end{figure}
	
\begin{figure}[!h]
		\centering
		\includegraphics[width=0.47\textwidth]{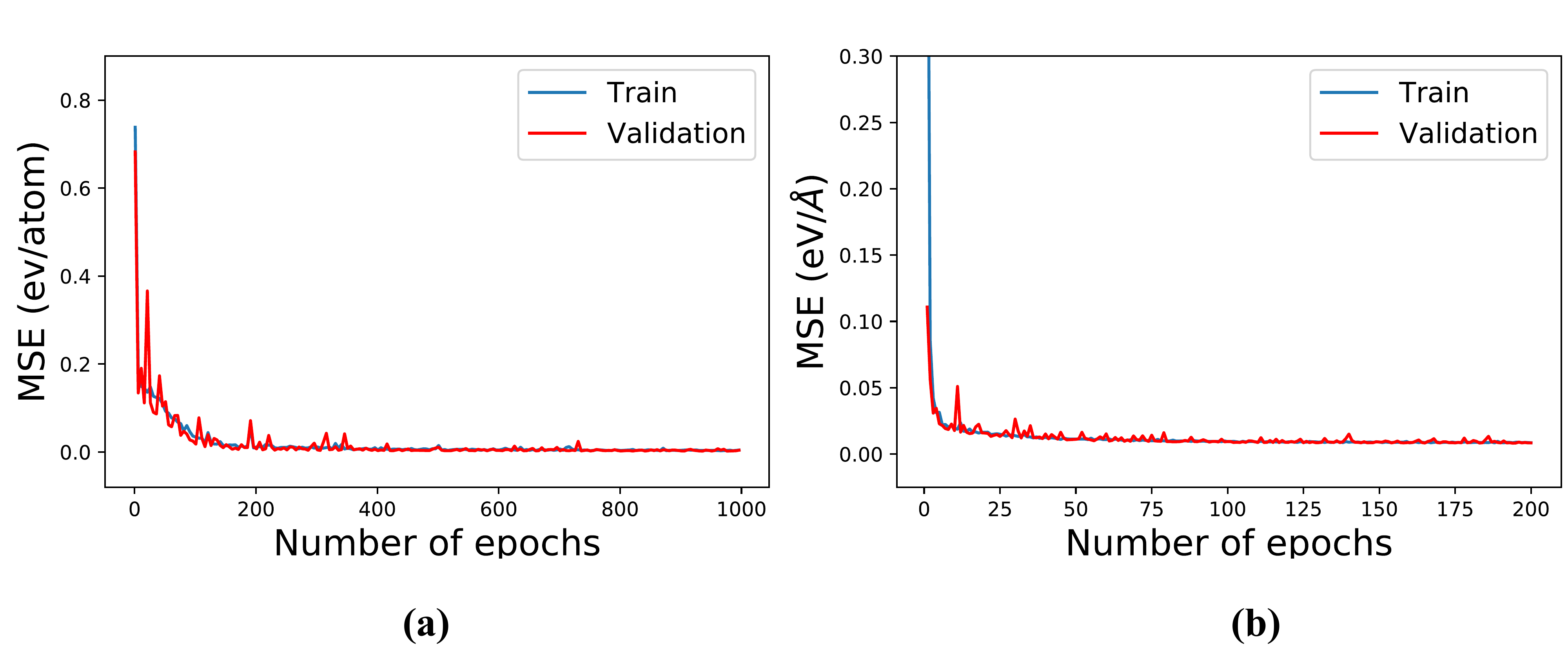}
		\caption{\label{fig:error}Statistical error metrics used to train model: energy model (a); force model 	(b)}
	\end{figure}

	We adopted 2D convolutional (Conv2D) layers implemented in the TensorFlow/Keras \cite{tensorflow2015-whitepaper} Library with $1 \times 1$ kernel to extract the embedding vector representing the pairwise terms, $\bm{a}_{ij}$. The number of embedding features was determined by the number of kernels (filters) of Conv2D layers, and we used three Conv2D layers with 128 neurons (filters). After a Conv2D layer, the input tensor of a structure transforms into a new tensor that has the same dimensions as the input tensor but the third being the number of filters of the Conv2D layer. To obtain the feature vectors of the local structures, we performed the sum over the third dimension the output tensor $\bm{x}_i = \sum_j \bm{a}_{ij}$, and as the result, we could obtain a matrix, ${X}$, with the rows being the feature vectors of the local structures. We then implemented the network for total energy by using another three 1D convolution (Conv1D) layers with 128 neurons and the kernel size of $1$ on the matrix ${X}$. In the final layer we used only one neuron to represent the atomic energy, and obtain the total energy by the sum over the atomic energies. The mean square error, $L_E = \frac{1}{m}\sum_{s=1}^m (E^{(s)} - \hat{E}^{(s)})^2 $ where $L_E$ is the loss function for energy model, $E^{(s)}$ and $\hat{E}^{(s)}$ are the energy of the $s^{th}$ structure by DFT and by our model, respectively, was used for the loss for training our network. Figure \ref{fig:model_energy} illustrates the general architecture of our model for predicting atomic energy, which is a collection of independent convolution neural networks (CNNs).  We employed batch normalization (also known as the batch norm) after each convolution layer to reduce the hidden unit values shifting around (covariance shift). We also use the regularization factor $l_2$ to apply penalty on layer parameters of the layer’s kernel and Dropout technique to control the overfitting, i.e., we add the norm2 of kernel's weights to the loss function that the network optimizes, and random connections between layers are freezed. We examined ``rectified linear unit" (Relu) \cite{Schmidhuber_2015}, Sigmoid, and Tanh activation functions for all hidden layers whereas a linear activation function is applied to the last layer. The Adam Optimizer \cite{Schmidhuber_2015} is used for optimizing the neural network. Fig \ref{fig:error} (a) illustrates the results of mean square error metric (MSE) for energy model. It can be seen that the learning curves decrease in both train and validation sets.
	
	Fig. \ref{fig:energy} (a) shows the comparison of energies estimated by our deep neural network and the DFT calculate one, which appear with a good agreement for most of the points on the scatter plot being in the straight line cutting through the origin with the slop of 1. Using Relu activation for reducing the computational cost, we obtained the root mean square error (RMSE) and mean absolute error (MAE) for the test set of 2.851 $meV/atom$ and 2.122 $meV/atom$, respectively. By applied Sigmoid activation, we obtained the RMSE and MAE for the test set of 2.664 $meV/atom$ and 1.566 $meV/atom$, respectively. It shows the significant improvement compare to the linear regression with MAE of 8.1 $meV/atom$ \cite{vnu_linear_regression} and the neural networks following Behler method \cite{BehlerPRL} using symmetry function with MAE of 6.1 $meV/atom$.

	\begin{table}[!h]
	\caption{\label{tab:error_energy} {Model performance for total energy prediction: the RMSE (meV/atom) and MAE (meV/atom)) estimated on the test set of $2\times 2 \times 2$ super cell.}}
	\centering
	\begin{tabular}{lp{1.9cm}p{2cm}p{2cm}}
	\hline\hline
                        & Activation & RMSE  & MAE \\
 	\hline
	\multirow{4}{*}{Conv1D} & Relu       & 2.851  & 2.122 \\
    	                    & Tanh       & 2.844  & 2.178 \\
        	                & Sigmoid    & 2.664  & 1.960 \\
            	            & Mix        & 2.155  & 1.566 \\
	\hline
	\multicolumn{2}{l}{Linear Gaussian}  & 4.530  & 3.348 \\
	\multicolumn{2}{l}{Linear Cosine}    & 8.691  & 6.348 \\
	\hline\hline
	\end{tabular}
	\end{table}

	As mentioned above, our model utilizes the idea of parameter sharing, and we use a single network for extracting hidden features of pairwise terms and another network for present partial energy for all atoms. Hence our model can be used for systems of any size. We examine our model for $3 \times 3 \times 3$ silicon system with 216 atoms. We randomized the silicon crystals by randomly displacing atomic positions and distorting the supercell, to construct 500 structures and performed DFT calculations for force and energy for these structures for reference data.  We employed the model with Relu activation to predict the total energy for this system. In comparison to DFT calculations, we obtained the RMSE of 8.67 $meV/atom$ for energy which indicates that our model can be applied for the larger system while trained with smaller systems. 
	
	We again emphasize that our model does not use the heuristic basic functions to encode the atom in a chemical environment for estimating its partial energy. Indeed, we employ a deep neural network to extract the hidden features of pairwise terms and stack them to derive the atomic feature vector. Fig. \ref{fig:energy} (b) show the typical basis functions extracted by our network using Relu activation. It is seen that our network can automatically learn to extract well-behaved functions basis functions very much like the Gaussian functions. To enrich the representation of basis functions, we examined the procedure of mixing Relu, Sigmoid, and Tanh activation function, in which three parallel Conv2D layers with the activation functions were employed to extract the embedding feature of the pairwise items, then these features were concatenated for the embedding vectors of the pairwise terms. Our experiments showed that this procedure was significantly approved with MAE of 1.566 $meV/atom$. The basis functions obtained by mix activation show more diversity than those by the Relu activation, as seen in Fig. \ref{fig:energy}(b) and (c). Hence it is a good consideration of other complicated systems.

	It is noteworthy that we can design the feature vectors of the pairwise terms to include the physics of the interactions such as many-body effects, Coulomb interactions, and other information such as the chemical environments of the pairwise terms. As a demonstration, we added the information of the local environment associated with the center atoms. We adopted the Coulom matrix of the atoms in the local chemical environment of an atom \cite{Rupps,Rupp_tutorial} to feature vectors of its interaction with neighboring atoms, i.e., we designed the feature vectors as $\bm{b}_{ij} = (\bm{r}_{ij}f_c(\bm{r}_{ij}), \frac{1}{\bm{r}_{ij}}f_c(\bm{r}_{ij}), \bm{s}_i)$, where $\bm{s}_i$ is the sorted eigenvalue of the Coulomb matrix. By integrating the Coulomb matrix information to the pairwise vectors, we can obtain the RMSE and MAE of {1.832 $meV/atom$} and {1.398 $meV/atom$} for test set of  $2 \times 2 \times 2$ system, and \mbox{{6.311} $meV/atom$} and {4.448 $meV/atom$} for $3 \times 3 \times 3$. For the more complex systems, we can design appropriate feature vectors for the pairwise terms. It is also noted that for some feature vectors of the pairwise terms, it is quite hard to calculate the atomic forces by the derivatives of energy model with respected to atomic coordinates.
	
	We also examined the ability of the machine learning model to the multicomponent system. Similar to the case of the Silicon system, we ultilized the primitive representation of a pair of atoms in the SiLi system.	We adopted the SiLi dataset of 5000 structures with DFT-calculated energy. We divided the dataset into the training set of 3750 structures and the test set of 1250 structures of the training set and test set, respectively. For multiple-component systems, a pair of atoms, $i$ and $j$, is represented by a vector: $(r_{ij}, 1/r_{ij}, Z_i, Z_j)$ and the pairwise term is expressed in term of: $\bm{b}_{ij} = (\bm{r}_{ij}f_c(r_{ij}), \frac{1}{\bm{r}_{ij}}f_c(r_{ij}), Z_i, Z_j)$. The matrix for representing an atom in the Si-Li system is modified to n $\times$ 3, where n is the number of neighboring atoms. We also used a cutoff radius of $8.0\mbox{\AA}$ for the SiLi system. The procedure and parameters of the deep neural network for initialization and training was the same as that of Si. 
	Our experiments showed that this procedure can be apply with RMSE and MAE of 4.117 $meV/atom$ and 3.236 $meV/atom$, respectively.

\subsection{Atomic forces}

	Our examination on the calculated force by obtaining the gradient of the total energy showed that the performance of the model on the force prediction is relatively poor, and the reason might attribute to the gradient vanishing of deep neural networks. And it might be hard to evaluate the derivatives when including the environment information with Coulomb matrix. To improve the prediction of force, we develop the deep neural network to represent the force exerted on an atom. To demonstrate our method, we also used the above $2 \times 2 \times 2$ silicon system with the dataset of 5000 structures and DFT-calculated atomic forces divided into the training set of 3750 structures and the test set of 1250 structures of the training set and test set, hence we had 240,000 force vectors associated with atoms in the training set and 80,000 of those for the test set.
	
	\begin{figure}[!h]
		\centering
		\includegraphics[width=0.49\textwidth]{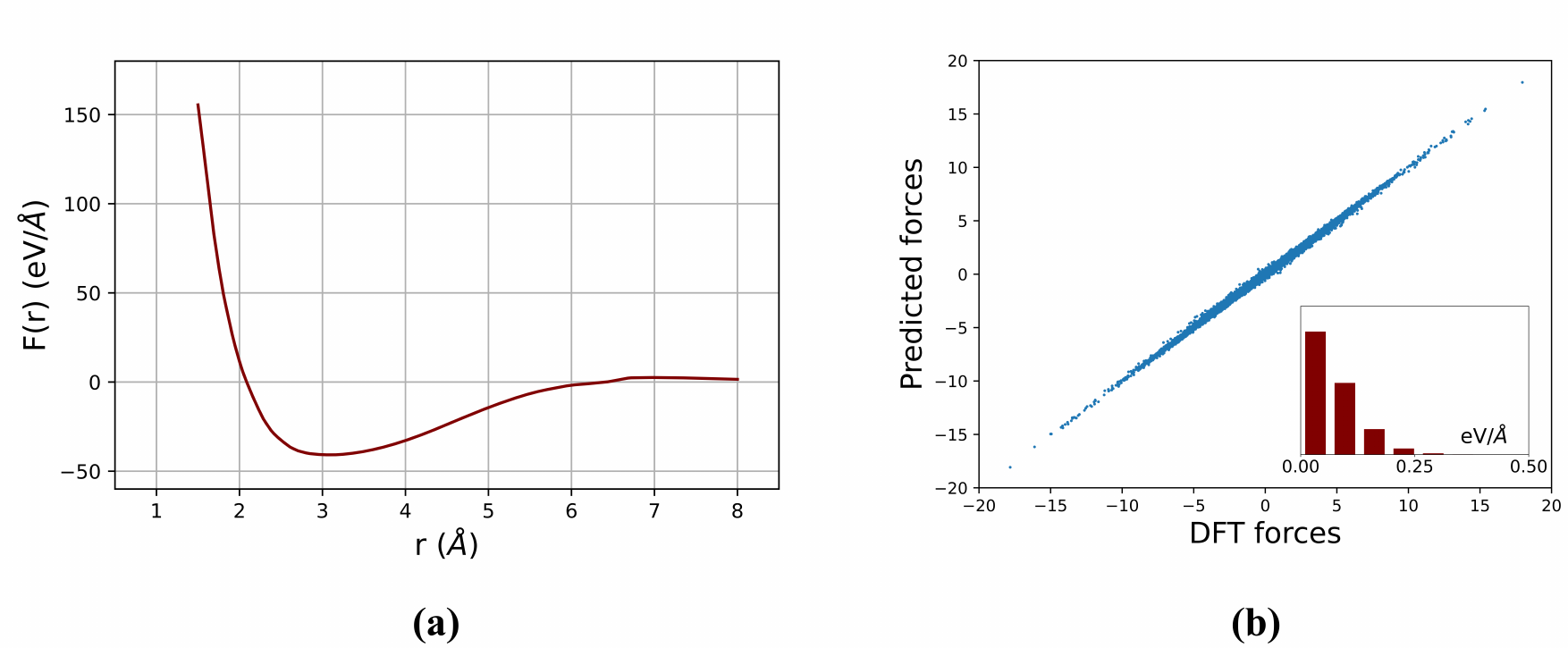}
		\caption{\label{fig:force} Shape of pairwise potential function (a) and Forces predicted by machine learning and DFT for silicon atoms in the test set (b)}
	\end{figure}
	
 	We used the cutoff radius of $8.0 \AA$ to determine the neighboring atoms of an atom, and the force exerted on it can be decomposed into the pairwise forces given by the neighboring atoms, as seen in Eq. \ref{eq:force}.     We also adopted the distance and its inverse as features for the pairwise term: $\bm{b}_{ij} = (\bm{r}_{ij}f_c(r_{ij}), \frac{1}{\bm{r}_{ij}}f_c(r_{ij})$. Hence, we obtain a matrix of ${n} \times 2$ for representing a silicon atom, where $n$ is the number of neighboring atoms, and the deep neural network with four hidden layers was used for the pairwise function, $\bm{F}_{ij}$. The network is shared by all pairs of silicon atoms. We also employed the padding technique to ensure all atoms having the same dimensional matrices, and in this silicon system, we found that $n = 150$, is enough. Unit vectors (in the Cartesian reference of the systems) for all pairs were also acquired, ${N}$ with the dimensions of $150 \times 3$.  Five 1D convolutional neural networks (Conv1D) layers with 128 neurons were used for extracting embedding features of pairwise terms and one Conv1D with one neuron for the pairwise function, $\bm{F}_{ij}$, as shown in Fig. \ref{fig:model_force}. The pairwise functions were then multiplied with the corresponding unit vectors for the pairwise forces, and we summed over the pairwise forces to obtain force for an atom, as described in Eq. \ref{eq:force}. We also investigated various activation functions for the feature extractors and network, including ``Rectified linear unit'' (ReLU), ``Tangent hyperbolic'' (Tanh), and ``Sigmoid''. We also used the mean square error of atomic forces to train our network:
 	\begin{equation}
 		L_F = \frac{1}{MN}\sum_{s=1}^M \sum_{i=1}^N\sum_{\alpha \in \{x, y, z\}}\left( F_{si}^\alpha - \hat{F}_{si}^\alpha \right)^2 \mbox{,}
 	\end{equation}
 	where $M$ and $N$ is the number of structures and number of atoms in the structures, respectively, and $ F_{si}^\alpha$ and $\hat{F}_{si}^\alpha$ are $\alpha$ component of force on atom $i$ of structure $s$ by DFT and our model prediction, respectively.
 	 Fig. \ref{fig:error} (b) shows the typical learning curves of our model with good learning behavior. 
	
	\begin{table}[!h]
		\caption{\label{tab:error_force} {Model performance for atomic force prediction: the RMSE ($eV/\AA$) and MAE ($eV/\AA$)) estimated on the test set of $2\times 2 \times 2$ super cell.}}
		\centering
		\begin{tabular}{lp{1.9cm}p{2cm}p{2cm}}
		\hline\hline
  	                      & Activation & RMSE  & MAE   \\
		 \hline
		\multirow{4}{*}{Conv1D} & Relu       & 0.096 & 0.076 \\
              		          & Tanh       & 0.098 & 0.074 \\
         		               & Sigmoid    & 0.095 & 0.073 \\
                		        & Mix        & 0.093 & 0.071 \\
		\hline
		\multicolumn{2}{l}{Linear Gaussian}  & 0.124 & 0.092 \\
		\multicolumn{2}{l}{Linear Cosine}    & 0.185 & 0.141\\
		\hline\hline
		\end{tabular}
	\end{table}

	Table. \ref{tab:error_force} summarizes the assessments of our model performance with RMSE and MAE obtained by Relu, Tanh, Sigmoid, and mixed activation functions, and Fig. \ref{fig:force}(a) shows the comparison of the atomic forces estimated by our deep model and the DFT forces in a good agreement. The results show that our model works well with all the activation functions with the RMSE less than 0.1 $eV/\AA$, and the mix activation outperforms over individual activation function and we can obtain the RMSE and MAE of {0.093} $eV/\AA$, and {0.071} $eV/\AA$. It shows the improvement compare to the linear model based on Gaussian and Cosine function with RMSE of 0.124 $eV/\AA$ and 0.185 $eV/\AA$, respectively. It is noteworthy that, in the above experiments, with only the primitive information of a pair of atoms (i.e., without any information about the chemical environment), the prediction of the atomic forces is surprisingly accurate, as seen in Fig. \ref{fig:force} (b). This result is even slightly more accurate than that in some of the previous works \cite{force_prediction_PRL,force_prediction_2}, and the forces obtained by the derivative of the energy model. This result implies that information of the chemical environment can be encoded in the NN by the training processes. An analysis of the output of the NN enables us to extract the hidden pairwise-force curve. Fig. \ref{fig:force}(a) shows the pairwise-force curve for the Si system. It is seen that the pairwise force is zero at the distance of nearly $2.3~\mbox{\AA}$ which is equilibrium Si-Si bond length in the silicon crystal.

	It is natural that the information of the environment also plays a role in the interaction of a pair of two atoms. Considering that the pairwise force between two atoms is a function of the two atoms, the distance between the two atoms, and information of the environment, for the Si system, we consider the following functional form for representing the pairwise forces: $\bm{F}_{ij} = F(\bm{r}_{ij}, 1/\bm{r}_{ij}, \bm{\bm{s}}_{i})$, where $\bm{\bm{s}}_{i}$ is a vector representing the environment between atoms $i$ and $j$. By adopting the information of the chemical environment and using Relu activation, we obtained the RMSE and MAE values for the randomized dataset of 0.0916 $eV/\AA$ and 0.0714 $eV/\AA$, respectively. The result indicates a significant improvement in accuracy when we include information of the chemical environment to represent the pairwise forces. We also examined the transferability of the NN forces trained by the $2\times2\times2$ silicon system, by comparing the NN forces and DFT forces of the $3\times3\times3$ silicon system. A test set with 500 structures of the $3\times3\times3$ supercell was used for the test. We obtained an RMSE of 0.095 $eV/\AA$ and MAE of $0.076~eV/\mbox{\AA}$ for the $3\times3\times3$ system. This result confirms that our model, which is trained with a small system, is transferable to large systems.
	
		\begin{figure}[!h]
			\centering
			\includegraphics[width=0.47\textwidth]{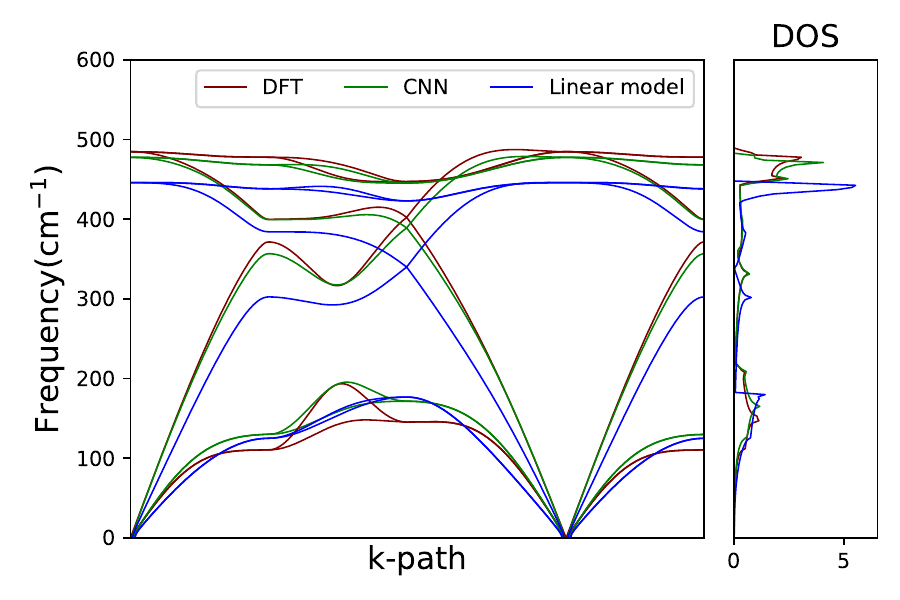}
			\caption{\label{fig:phonon}Phonon dispersion of diamond Si with a $2\times2\times2$ supercell (left) and phonon density of state (DOS) (right): calculated by DFT (red) and calculated by NN (green).}
		\end{figure}

	We considered a SiLi dataset consisting of 5,000 structures and 1,050,000 force components of all atoms. We randomly selected 262,500 forces as the test set and the remaining 787,500 forces served as the training set. Similar to the pure Si system, improving the performance of our model may require information of the chemical environment to represent the pairwise force. We propose the following functional form to represent this force in multicomponent systems: $\bm{F}_{ij} = F(\bm{r}_{ij}, 1/\bm{r}_{ij}, Z_i, Z_j, \bm{s}_i)$, where $\bm{s}_i$ is a vector representing the environment between atom $i$ and $j$. We obtained an RMSE of $0.118 eV/\AA$ and MAE of $0.086~eV/\mbox{\AA}$ for the SiLi system.
	
	We investigated the applicability of the latest NN to the calculation of phonon dispersion for Si with a $2\times2\times2$ supercell. The phonon dispersions were calculated by using the supercell approach \cite{phonon_model} implemented in Phonopy code \cite{phonopy}. Figure \ref{fig:phonon}, which shows the phonon dispersion for diamond Si using ML and DFT calculations, shows that the phonon dispersions obtained with these two methods are in good agreement. The slight difference indicates the limit of NN with respect to the prediction of small forces.

\subsection{Thermodynamic properties}
	In this section, we concentrated on the thermal properties, thermal expansion, and mechanical properties of silicon systems at constant pressure by quasi-harmonic approximation (QHA) \cite{qha_PhysRevB.81.174301}. QHA estimates the thermal expansion of a crystal by combining the vibrational energy of the phonons to total free energy \cite{thermal_PhysRev.112.136}. This allows volumetric expansion can be expressed in terms of temperature, $V (T )$, which is necessary to estimate the contribution of thermal to the elastic constants.
	\begin{figure}[!h]
			\centering
			\includegraphics[width=0.47\textwidth]{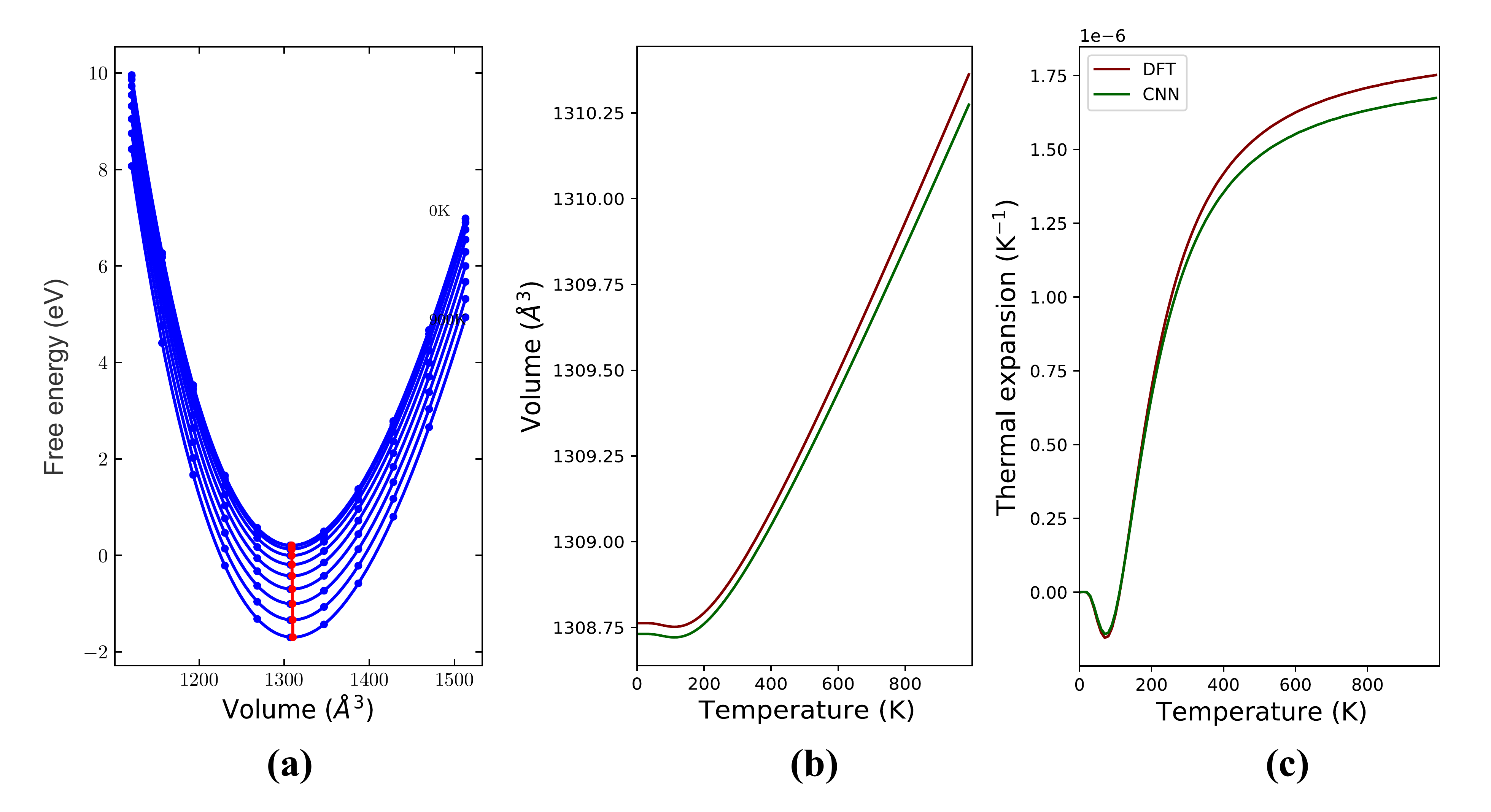}
			\caption{\label{fig:thermal} Thermal properties}
	\end{figure}
	To represent the volume as a function of temperature, we applied a set of strains to collect corresponded structures. A range of temperatures from 0 to 1000 Kelvin were also chosen to expand the reference volume. The result is shown in Fig \ref{fig:thermal} (a). The set of points corresponding to each temperature is then fit using polynomial interpolation. The minimum of each curve, shown by red points is corresponded to the total free energy density at zero pressure.
	Fig \ref{fig:thermal}(b) illustrates the volumetric expansion as a function of the temperature  $V(T )$ of the silicon system. It shows that the result obtained by the deep neural network is accurately comparable with DFT. The calculated thermal expansion coefficient of silicon system by deep neural network compared to DFT calculation is shown in Fig \ref{fig:thermal}(c). Silicon is also recognized to exhibit negative thermal expansion (NTE) at low temperatures \cite{Kim1992}. The computed thermal expansion coefficient of silicon displayed in Fig \ref{fig:thermal}(c) exhibits NTE at low temperature further proving the validity and implementation of the deep neural network by QHA.
	
	\begin{figure}[!h]
			\centering
			\includegraphics[width=0.47\textwidth]{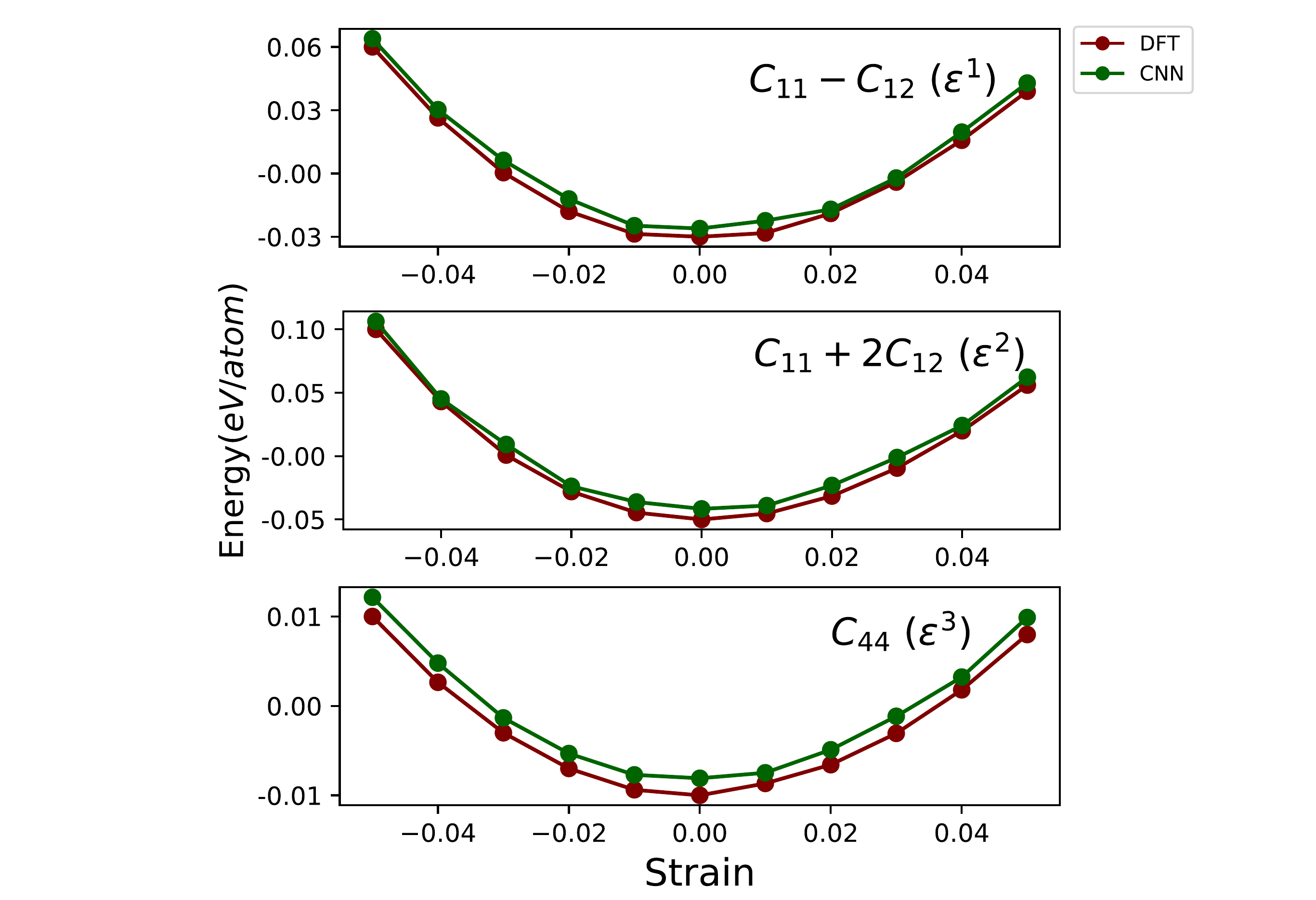}
			\caption{\label{fig:elastic} Comparison of different elastic constants derived using the energy vs strain trends which computed by DFT and NN models.}
	\end{figure}
	
	\begin{table}[!h]
		\caption{\label{tab:strain} {Parametrization of the strains applied to calculate the internal energy, $\gamma = \pm 0.01n$ where $n = \overline{0,5}$.}}
		\centering
		\begin{tabular}{lll}
		\hline\hline
  	                       Strain & Parameters (unlisted $\varepsilon_{ij} = 0$) & $\Delta E/V_0$ to $O(\gamma ^2)$  \\
		\hline
		$\varepsilon^1$  &  $\varepsilon_{11} = \varepsilon_{22} = \gamma, \varepsilon_{33} = -2\gamma$  & $3(C_{11} - C_{12})\gamma^2$ \\
		$\varepsilon^2$  &  $\varepsilon_{11} = \varepsilon_{22} = \varepsilon_{33} = \gamma$ & $3/2(C_{11} + 2C_{12})\gamma^2$  \\
		$\varepsilon^3$ &  $\varepsilon_{12} = \varepsilon_{21} = \gamma/2$ & $(C_{44} \gamma^2)/2$\\
		\hline\hline
		\end{tabular}
	\end{table}
	
	Elastic constants play an important role in determining the mechanical properties of material which can be estimated by calculating the total energy as a function of appropriate lattice deformations \cite{elastic_PhysRevB.68.184108, elastic_DING2012555}. We calculated the corresponding variations in the total energy by applying small strains to the equilibrium unit cell. In this case for cubic material, there are three independent elastic constants $C_{11}, C_{12}$ and $C_{44}$ which can be determined by straining the lattice vectors according to $\Vec{\Gamma}' = (I + \varepsilon)(\Vec{\Gamma})$ as described in Data section. The internal energy of a crystal applied strain $\varepsilon$ listed in Table. \ref{tab:strain} can be expressed in terms of the strain tensor:
 		\begin{equation}
 		E(V,\{{\varepsilon_i}\}) = E(V,0) + \frac{V_0}{2}\sum_{ij}C_{ij}\varepsilon_i\varepsilon_j + ...
 		\end{equation}
	where $V_0$ is the volume of undeformed structure corresponding to energy $E(V_0,0)$, $C_{ij}$ is elastic constant tensor, and strain tensor $\varepsilon = \{\varepsilon_i, \varepsilon_j,...\}$ are given in Voigt notation.
	Fig \ref{fig:elastic} illustrates the total energy - strain curves of silicon system obtained by DFT and CNN. It can be seen that the elastic constants are in reasonable agreement with the DFT calculation results.

\section{Conclusions}
	We propose a full-stack model using multi-layers convolutional neural networks to represent forces and energy to facilitate the MD simulation. In our model the deep NN are designed to transfer the information of two interaction atoms (the pairwise interactions) and  the chemical environment into embedding their features for predicting atomic force and energy. We demonstrated that our model can well reproduce the DFT-calculated forces and energy of silicon systems, and the accuracy can be improved by integrating the information of the chemical environment to the pairwise interactions. We obtained RMSE and MAE for the atomic force of approximately 0.07 $eV/\AA$ and 0.09 $eV/\AA$, respectively, and those for the atomic energy of 1.832 $meV/atom$ and 1.398 $meV/atom$, respectively. We also revealed that our trained model with $2\times2\times2$ systems can be applied to estimate atomic forces and energy for $3\times3\times3$ systems comparable to those obtained by DFT calculations. Furthermore, the performance of our model is satisfactory for predicting the force and energy of not only Silicon systems but also multi-component systems. The force and energy models were also utilized to calculate phonon dispersion and thermodynamic properties of silicon which illustrated in a good agreement with DFT.  
	
\section*{Acknowledgments}
	
	This research was funded by the Vietnam National Foundation for Science and Technology Development (NAFOSTED) under grant number 103.01-2019.30. Nguyen Van Quyen was funded by Vingroup Joint Stock Company and supported by the Domestic Master/ PhD Scholarship Programme of Vingroup Innovation Foundation (VINIF), Vingroup Big Data Institute (VINBIGDATA), code VINIF.2020.ThS.17
\section*{References}
\bibliography{main}

\end{document}